\documentclass[journal=jctcce,layout=twocolumn,manuscript=article]{achemso}
\setkeys{acs}{articletitle = true}
\usepackage{amsmath}
\usepackage{bm}
\usepackage{dcolumn}
\newcolumntype{d}[1]{D{.}{.}{#1}}
\usepackage{graphicx}
\usepackage{multirow}
\usepackage{lineno}
\usepackage{txfonts}
\usepackage{booktabs}
\usepackage{subcaption}
\usepackage{gensymb}
\usepackage{siunitx}
\usepackage{adjustbox}
\usepackage[usenames, dvipsnames]{color}
\usepackage{wrapfig} 
\usepackage{bm}
\usepackage{mathrsfs}

\title{Good rates from bad coordinates: the exponential average time-dependent rate approach} 

\let\oldmaketitle\maketitle
\let\maketitle\relax
\usepackage[fontsize=11pt]{fontsize}
\author{Nicodemo Mazzaferro}
\affiliation{Department of Chemistry, New York University, NY, 10003, USA}

\author{Subarna Sasmal}
\affiliation{Department of Chemistry, New York University, NY, 10003, USA}

\author{Pilar Cossio}
\email{pcossio@flatironinstitute.org}
\affiliation{Center for Computational Mathematics, Flatiron Institute, New York, 10010, USA}
\altaffiliation{Center for Computational Biology, Flatiron Institute, New York, 10010, USA}

\author{Glen M. Hocky}
\email{hockyg@nyu.edu}
\affiliation{Department of Chemistry, New York University, NY, 10003, USA}
\altaffiliation{Simons Center for Computational Physical Chemistry, New York University, NY, 10003, USA}

\begin{document}
\twocolumn[
\begin{@twocolumnfalse}
\oldmaketitle
\begin{abstract}
Our ability to calculate rates of biochemical processes using molecular dynamics simulations is severely limited by the fact that the time scales for reactions, or changes in conformational state, scale exponentially with the relevant free-energy barriers. 
In this work, we improve upon a recently proposed rate estimator that allows us to predict transition times with molecular dynamics simulations biased to rapidly explore one or several collective variables. 
This approach relies on the idea that not all bias goes into promoting transitions, and along with the rate, it estimates a concomitant scale factor for the bias termed the collective variable biasing efficiency $\gamma$.
First, we demonstrate mathematically that our new formulation allows us to derive the commonly used Infrequent Metadynamics (iMetaD) estimator when using a perfect collective variable, $\gamma=1$.
After testing it on a model potential, we then study the unfolding behavior of a previously well characterized coarse-grained protein, which is sufficiently complex that we can choose many different collective variables to bias, but which is sufficiently simple that we are able to compute the unbiased rate directly.
For this system, we demonstrate that our new Exponential Average Time-Dependent Rate (EATR) estimator converges to the true rate more rapidly as a function of bias deposition time than does the previous iMetaD approach, even for bias deposition times that are short. 
We also show that the $\gamma$ parameter can serve as a good metric for assessing the quality of the biasing coordinate.
Finally, we demonstrate that the approach works when combining multiple less-than-optimal bias coordinates.
\end{abstract}
\end{@twocolumnfalse}]

\maketitle 

\section{Introduction}
A major challenge in biomolecular simulation is to be able to accurately assess the transition rates (inverse of the mean residence time in a state) of complex processes, including conformational transitions and the binding/unbinding of macromolecules and their ligands.
Processes of interest often involve rare events, where the system spends a large amount of time in a metastable state and rarely transitions to another relevant one, so the transition path time is typically orders of magnitude shorter than the time spent in either state \cite{chandler1998barrier}.
Because of this, extracting rates directly from unbiased simulation is out of reach for all but the simplest of systems.

Numerous methodologies have been developed to accelerate rare conformational transitions, with the primary purpose being to compute ensemble-averaged observables \cite{henin2022enhanced,tuckerman2023statistical}.
A major subclass of such methods operate by adding an additional biasing potential to the system's Hamiltonian, usually in terms of a small set of collective variables (CVs) which are believed or determined to be good descriptors of states of interest, or the path between them \cite{henin2022enhanced,tuckerman2023statistical}.
Common examples of such methods include Umbrella Sampling, Adaptive Bias Force, Metadynamics (MetaD), Variationally Enhanced Sampling, and On-the-fly Probability Enhanced Sampling, among others \cite{torrie1977nonphysical,kastner2011umbrella,darve2001calculating,comer2015adaptive,metad,wtmetad,bussi2020using,valsson2014variational,invernizzi2020rethinking}.
All of these methods pay the price of distorting the system's dynamics to obtain 
a much more rapid estimate of the underlying free-energy landscape as a function of the CVs.

Most methods that tackle the problem of computing rates of rare transitions seek to generate a set of unbiased trajectories, either through combining direct sampling of many short trajectories from different starting points \cite{MSM_method,thiede2019galerkin} as in Markov State Modeling, through Monte Carlo in trajectory space as in Transition Path Sampling\cite{TPS_method}, or by generating trajectories that progress in a particular coordinate as in Forward Flux Sampling \cite{allen2009forward}, Steered Transition Path Sampling \cite{guttenberg2012steered}, Weighted Ensemble \cite{zuckerman2017weighted}, Transition Interface Sampling \cite{TIS_method}, and Milestoning \cite{elber2020milestoning}.
However, these methods are computationally expensive, and some scale poorly with system size, making them challenging to apply for the complex biophysical problems we are interested in studying, such as finding the timescale for protein-drug unbinding\cite{casasnovas2017unbinding}, for the RBD opening of the SARS-CoV-2 Spike protein \cite{wieczor2023omicron} or for the unbinding of cytoskeletal adhesion proteins under force \cite{gomez2021molecular,pena2022assessing,mukadum2023molecular}.

As such, we are interested in approaches that build on CV biasing methods, which have been used to probe conformational transitions with sufficient computational efficiency even for relatively complex biological assemblies. 
The challenge already mentioned is that these methods alter the dynamics, which prevents any obvious solutions to inferring the unbiased time scales of events.
However, starting with the Hyperdynamics method of Voter, it was shown that the first passage time of rare events could be approximately predicted using biased simulations if bias is not applied during the actual crossing through the transition state, by formulating an ansatz for how the bias accelerates time  \cite{voter1997hyperdynamics,voter1997method}.
This approach was originally developed using a time-independent potential defined on the whole system of interest, but later in the Infrequent Metadynamics (iMetaD) approach the same ideas were extended to CV biasing. 
MetaD \cite{metad} works by updating an external bias with a Gaussian centered at the current position in CV space every $\Delta$ time steps (see Sec.~\ref{sec:wtmetad} for details). iMetaD solves the problem of not biasing the transition over the barrier by only rarely updating the bias potential, such that it is unlikely to add bias on a high barrier during a fast crossing. iMetaD also introduces an additional approximation since the system is experiencing a time-dependent bias rather than a static one. 
The difficulty of avoiding adding bias during barrier crossings can also be mitigated by MetaD variants that only add bias within a region or up to a certain energy level \cite{mcgovern2013boundary,dama2015exploring}, which is now particularly easy to implement in the OPES variant of MetaD \cite{ray2022rare}.
iMetaD and similar approaches have now been used and benchmarked for many different problems, especially for protein-ligand unbinding problems, as detailed in Ref.~\citenum{ray2023kinetics}. 

To extract the transition rate, these methods assume that a `good' CV is used, and validate the rate estimates using a Kolmogorov-Smirnov (KS) test between the empirical and theoretical survival distributions. Unfortunately, for large and complex transitions, this might not be the case because finding a good CV is challenging. Moreover, CV quality indicators, such as the committor \cite{bolhuis2002transition}, are expensive or intractable to compute.
The Kramers time-dependent rate (KTR) method\cite{ktr} was recently developed to extract transition rates from biased simulations, such as those used for iMetaD, but with much less sensitivity to CV choice. It introduced a new parameter $\gamma$, called the CV biasing efficiency, to be extracted from the biased simulations which scales the bias. In that work, it was shown that $\gamma$ had a lower value for a poor CV in a simple 2D double-well potential, and as such it was assumed to relate to the CV quality \cite{ktr}. However, this has not been systematically demonstrated, and KTR has not been benchmarked on a problem where many CVs could be tested. Moreover, a direct connection between the KTR and iMetaD estimators has not been established.

In this work, we introduce a more general framework for computing rates from time-dependent biasing protocols, which allows us to treat the iMetaD and KTR estimators on the same footing.
We then use this framework to propose a revision to KTR termed the Exponential Average Time-dependent Rate (EATR) method that bridges the two approaches.
The EATR approach is shown to give the correct Kramers' rate when $\gamma=1$ for an idealized 1D potential.
Then, we use a G\=o-protein system as a model to show how the prediction of rates depends on the choice of bias coordinate, and compare EATR's results to the true intrinsic rate. Importantly, we find that $\gamma$ correlates with the intuition of CV quality. We find that for the poor biasing coordinates, the original KTR and EATR results are comparable and they enable an accurate recovery of the unbiased rates. Surprisingly, this is often true even in the frequent-biasing regime. The paper is organized as follows. First, we present a general theory for rate calculations from time-dependent biased simulations. We relate it to iMetaD and KTR, and then formulate the EATR approach. 
Then, we show results for a 1D overdamped Langevin dynamics, and for the unfolding process of protein G.
We also show that the method can be extended beyond one biasing coordinate, presenting accurate results on protein G unfolding when biasing two CVs simultaneously. 
We end with conclusions and future perspectives of the work.

\section{Theory}
\label{sec:theory}

\subsection{Transition rate for rare events}

The rare event problem constitutes the stochastic crossing of a single free-energy barrier, where typically the waiting time to cross the barrier is much longer than the transition time over it. 
For a high barrier, the survival function $S(t)$, which is the probability of a transition not occurring before time $t$, is given by an exponential distribution characterized by a single transition rate constant $k_0$,
\begin{equation}
    S(t)=e^{-k_0 t}~.
    \label{eq:poisson_survival}
\end{equation}
Note that this survival probability is related to the probability of a transition occurring \textit{at} time $t$ via 
\begin{equation}
p(t)=-\frac{dS(t)}{dt}~,
\label{eq:derivative_probability}
\end{equation}
and it is also related to the cumulative distribution function (CDF), the probability that a transition occurred \textit{by} time $t$, 
\begin{equation}
\mathrm{CDF}(t)=1-S(t)~.
\label{eq:cdf}
\end{equation}

For Brownian dynamics, Kramers' rate theory\cite{orig_kramers,szabo_kramers,hanggi1990reaction} can be used with several approximations to calculate the barrier crossing rate, $k_0$, from the bottom of a well in a potential $U(x)$ with a single high barrier, 
\begin{equation}
    k_0= D \int_{\mathrm{well}} e^{-\beta U(x)} dx \int_{\mathrm{barrier}}e^{\beta U(x)} dx~,
    \label{eq:kramers}
\end{equation}
where $D$ is the diffusion coefficient. 

However, for most systems of interest, the diffusion coefficient and underlying potential (or free-energy landscape) are not known and, therefore, one cannot directly use Eq. \ref{eq:kramers} to estimate the rate. Instead, the transition rate can be calculated using the survival function and a set of simulations $i=1,\dots,N$ where $M\leq N$ have crossed the barrier and $N-M$ have not. Let $t_i$ be the time the $i$-th simulation crossed the barrier, and $t_i=T_i$ the total simulation time for simulations $i=M+1,\dots,N$. For right censored transition times, the likelihood is given by
\begin{equation}
   \mathscr{L}=\prod_{i=1}^{M} p(t_i)\,\, \prod_{i=M+1}^{N} S(T_i)~,
   \label{eq:likeli}
\end{equation}
which is the product of the probabilities of the transitions occurring at times $\{t_i\}_{i=1}^{M}$ and the probabilities of not transitioning before the times $\{T_i\}_{i=M+1}^{N}$ for $N-M$ simulations \cite{likelihood_function}.  

An estimate for the transition rate can be obtained by substituting Eq.~\ref{eq:poisson_survival} and Eq.~\ref{eq:derivative_probability} into Eq. \ref{eq:likeli} and maximizing the logarithm of the likelihood with respect to $k_0$, 
\begin{equation}
    k_0^*=\frac{M}{\sum_{i=1}^N t_i}~. 
    \label{eq:rate_poisson}
\end{equation}
Note that the summation in the denominator takes into account the simulations that did not transition. When all simulations have crossed the barrier, Eq. \ref{eq:rate_poisson} reduces to the inverse of the average barrier-crossing time, $k_0=\left< t_i\right>^{-1}$ where $\left<  \cdot \right>$ denotes the average over the simulations.  

The transition rate can also be calculated by fitting the CDF (Eq. \ref{eq:cdf}). 
To do so for the same set of simulations $i=1,\dots,N$, we  construct an empirical CDF $h(t_i)$ which is the number of simulations that have transitioned before $t_i$ over the total number of simulations. The theoretical CDF can then be fit to the empirical CDF with a least-squares method by optimizing $k_0$.

\subsection{Expression for the general time-dependent rate}

For time-dependent biased simulations, such as a MetaD simulation, the transition rate is no longer a constant, and so as in Ref.~\citenum{ktr}, we consider how this time-varying rate could be taken into account. As a general expression, we can write the time-dependent rate as the unbiased rate $k_0$ scaled by a function of time $f(t)$ such that
\begin{equation}
    k(t)=k_0\,f(t)~.
    \label{gen_tdr}
\end{equation} 
We consider $f(t)$ to account for the effect of the time-dependence of the bias on the rate. In the following, we will relate it to different rate-extraction theories, such as iMetaD. 

Assuming a quasi-adiabatic biasing process, the survival function for the general expression of the time-dependent rate is given by  
\begin{equation}
    S(t)=e^{-\int_0^t k(t')\, dt'} = e^{-k_0\int_0^t f(t')\,dt'}~,
    \label{tdr_poisson_survival}
\end{equation}
which can be used in Eq.~\ref{eq:cdf} for fitting to an empirical CDF. 
Substituting this survival probability for the time-dependent rate into Eq.~\ref{eq:likeli} results in a general likelihood given by
\begin{equation}
    \mathscr{L}=\prod_{i=1}^{M}k_0f(t_i)e^{-k_0\int_0^{t_i} f(t') dt'} \prod_{i=M+1}^{N}e^{-k_0\int_0^{T_i} f(t') dt'}~,
\end{equation}
which can be simplified by taking its logarithm
\begin{equation}
    \log(\mathscr{L})= M\log( k_0) + \sum_{i=1}^{M} \log( f(t_i)) - k_0\sum_{i=1}^N \int_0^{t_i} f(t') dt'~. 
    \label{gen_log_likelihood}
\end{equation}
Similarly to the unbiased case, we can maximize this expression with respect to $k_0$, to obtain $k^*_0$ as an estimator for the true rate,
\begin{equation}
    k^*_0=\frac{M}{\sum_{i=1}^N\int_0^{t_i}f(t')dt'}~.
    \label{eq:gen_rate_estimator}
\end{equation}

\subsection{Relation to iMetaD}

In the hyperdynamics method \cite{hypermd,voter1997hyperdynamics}, the rate from transition state theory is scaled by the acceleration factor $\alpha=\left<e^{\beta V(x)}\right>_X$ (note that here we use the subscript $X$ to denote a configurational average and unlabeled brackets to represent an average over separate trajectories) where $V(x)$ is a fixed bias function added to the system's Hamiltonian as a function of the system's full coordinates.
This $\alpha$ arises by considering the average effect in many individual trajectories whose time is dilated by a factor $e^{\beta V_i(t)}$, where $V_i(t)$ is the bias experienced by the system during simulation $i$ at time $t$. 
Hyperdynamics then corresponds to a rate scaling function of the form
\begin{equation}
    f(t)=\left<e^{\beta V(x)}\right>_X=\alpha,
    \label{imetad_f}
\end{equation}
and replacing it in Eq.~\ref{tdr_poisson_survival} results in the survival probability
\begin{equation}
    S(t)=e^{-k_0\alpha t}~.
    \label{imetad_surv}
\end{equation}
Using this expression and assuming all simulations transitioned ($M=N$), the likelihood maximization of the rate gives
\begin{equation}
    k^*_0=\frac{1}{\alpha \left< t_i \right>}~.
    \label{hyperfold_rate}
\end{equation}

In iMetaD, the form of the bias is also changing in time along with the configuration of the system in a history-dependent manner.
Therefore, in iMetaD an acceleration factor for each simulation is approximated as a time average over that simulation instead of calculating it as a configurational average, $\alpha_i=\frac{1}{t_i}\int_{0}^{t_i} e^{\beta V_i(t')}dt'$. In the Supplementary Information, we show that if we use this expression in our time-dependent rate formulation, we recover the standard rescaled rate theory found for iMetaD,
\begin{equation}
    k^*_0=\frac{1}{\left< \alpha_i t_i \right>}~.
    \label{eq:imetad_rate}
\end{equation}
We note that this result is derived using the LM approach for the case where all simulations have transitioned. 

In Ref.~\citenum{imetadcdf}, it was shown that directly fitting the theoretical CDF (obtained from Eq.~\ref{imetad_surv}) is less sensitive to outliers in the tail of the distribution. The KS test can be used to assess whether the transition distribution is well-described by the theoretical CDF (see Sec.~\ref{sec:ks}).
Recently, it was also suggested that the short time information from the CDF can be fit to get a more robust estimate of the rate \cite{blumer2024short}.
We note that the results from likelihood maximization (LM) and CDF fitting need not coincide, as we will describe below.

\subsection{Kramers time-dependent rate and the CV biasing efficiency}

Most of the transition-rate methods for biased simulations, such as those described above, assume that bias is applied along a perfect CV, where all added bias accelerates the barrier crossing event. 
However, for large biomolecular systems, choosing \textit{a priori} a perfect CV for accelerating transitions to another targeted state is almost impossible. In practice, the bias is applied along non-ideal CVs, which insert bias along useless directions that are not aligned with the transition path. 

To overcome this issue, the KTR theory \cite{ktr} introduces a parameter, $\gamma\in[0,1]$, to account for the efficiency of the biased CVs. In addition to the unbiased rate, $\gamma$ will also be estimated from the simulation transition times, and it will inform about the quality of the CV with $\gamma\rightarrow 0$ reflecting poor CVs and $\gamma\sim 1$ good ones. 
In the KTR approach as previously implemented, the efficiency of CVs is accounted for by defining the scaling function as  
\begin{equation}
    f(t)=e^{\beta \gamma \left<\max V_i(t)\right> },
    \label{ktr_f}
\end{equation}
where $\left<\max V_i(t)\right>$ is the maximum bias applied at any point up to time $t$ in simulation $i$, averaged over all simulations (denoted $V_{MB}(t)$ in Ref.~\citenum{ktr}). This form of treating the biasing potential was inspired by rate-calculation methods developed for force-spectroscopy\cite{Dudko2006,Cossio2016}, where the barrier is reduced due to the external force, and therefore, by using $\left<\max V_i(t)\right>$ it was assumed that the bias only affects the barrier height. 
Inserting Eq. \ref{ktr_f} into Eq. \ref{tdr_poisson_survival} gives
\begin{equation}
    S(t)=e^{-k_0\int_0^t  e^{\beta \gamma\left<\max V_i(t)\right> } dt'}~,
    \label{ktr_surv}
\end{equation}
which can be used directly in a CDF fit. 
Substituting this expression into the log-likelihood from Eq. \ref{gen_log_likelihood}, and maximizing it with respect to $k_0$ results in a $\gamma$-dependent expression for the unbiased rate
\begin{equation}
    k^*_0(\gamma)=\frac{M}{\sum_{i=1}^N\int_0^{t_i}e^{\beta \gamma \left<\max V_i(t)\right>}dt'}~.
    \label{ktr_rate}
\end{equation}
 To obtain the maximum likelihood estimate for both $\gamma$ and $k_0$, Eq. \ref{ktr_rate} is substituted back into the log-likelihood function and it is maximized numerically with respect to $\gamma$. 

\subsection{Exponential average time-dependent rate (EATR)}

While rates estimated by the KTR approach are accurate (as shown in Ref.~\citenum{ktr} and in the following sections), we show below in Sec.~\ref{sec:1d} that it has the undesirable property that it does not agree with the iMetaD estimator when $\gamma=1$, whereas we expect the iMetaD estimator to be correct for an ideal coordinate with a very high barrier and slow deposition time. The reason for this discrepancy is the way in which the average effect of the bias is defined---for iMetaD $e^{\beta V_i(t)}$ is averaged, whereas for KTR the maximum of the biasing potential is averaged.

To unify the two theories, we propose the following modification to the KTR method, which will have the desired property of producing the same rates as iMetaD in the case where $\gamma=1$.  
To do so, we introduce the scaling function
\begin{equation}
    f(t)=\left<e^{\beta \gamma V_i(t)}\right>~.
\end{equation}
This gives the survival probability
\begin{equation}
    S(t)=e^{-k_0 \int_0^t \langle e^{\beta \gamma V_{i}(t') } \rangle dt'}~.
    \label{eq:survival_avgvi}
\end{equation}
Substituting this expression in Eq. \ref{gen_log_likelihood} results in a log-likelihood of the form
\begin{equation}
\begin{aligned}
     \log\mathscr{L}=& M\log k_0 + \sum_{i=1}^{M} \log \langle e^{\beta \gamma V_i(t_i)} \rangle \\
     &- k_0\sum_{i=1}^N \int_0^{t_i} \langle e^{\beta \gamma V_i(t')} \rangle dt'~.
     \end{aligned}
\end{equation}

In the case where all simulations transition ($M=N$), the optimal unbiased $k_0$ as a function of $\gamma$ is given by,
\begin{equation}
k_0^*(\gamma) = \frac{1} {\left \langle  \int_0^{t_i} e^{\beta \gamma V_{i}(t')} dt' \right \rangle}~,
\end{equation}
where we have benefited from the idempotence of averages to rewrite the average of an average as a single average. 
Observing that when $\gamma=1$, the term within brackets in the denominator is equivalent to $t_i \alpha_i$, this estimator is then identical to the standard iMetaD estimator in Eq~\ref{eq:imetad_rate}. Similarly as with the KTR, we can replace this expression into the log-likelihood and numerical maximize it with respect to $\gamma$ to obtain estimates for both the unbiased rate and the efficiency of CVs.

Importantly, we note that Eq.~\ref{eq:survival_avgvi} also provides us the option to numerically fit the biased empirical CDF to find the best values of $k_0$ and $\gamma$. As initial guesses, we use the LM estimates for $k_0$ and $\gamma$, and optimize using the Levenberg-Marquardt algorithm implemented in the SciPy Python package\cite{scipy,leastsq_fitting} to fit the empirical CDF to the theoretical CDF obtained from Eq.~\ref{eq:survival_avgvi}. The same can be done for the KTR method using the theoretical CDF from Eq.~\ref{ktr_surv}.

\section{Results and Discussion}

\subsection{Benchmarking on a 1D Potential}
\label{sec:1d}
\begin{figure}[ht]
\includegraphics[width=\columnwidth]{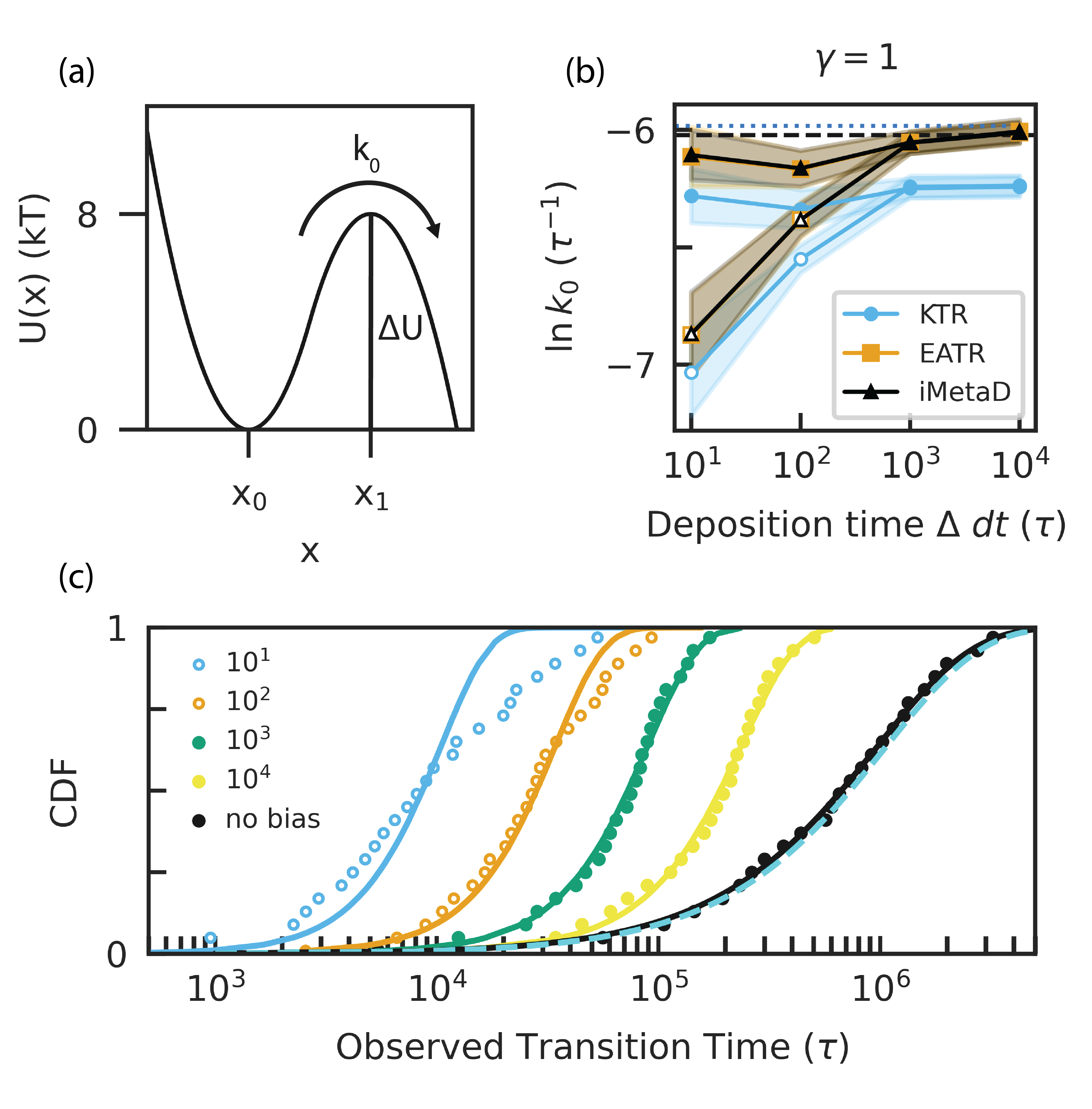}
\caption{\label{fig:results_1d} (a) The potential energy profile of the 1D matched-harmonic potential from Eq. \ref{eq:matchedharmonics}. (b) The unbiased rate from Kramers' rate theory and from unbiased simulations are shown as the dashed black line and the dotted blue line, respectively. We compare these with the predicted rates from iMetaD, and from the KTR and EATR methods by asserting $\gamma=1$ as a function of the bias-deposition time ($\Delta$ dt). Maximum likelihood estimates are represented by open symbols while CDF-fit estimates are filled symbols. Error bars are from a bootstrap analysis as described in Sec.~\ref{sec:bootstrap}. (c) The empirical CDFs of observed transition times over the barrier are shown with their EATR-CDF fits; different bias deposition times are indicated in the caption with curves showing fastest to slowest appearing from left to right. Fits that fail the KS test are represented with open symbols. The unbiased empirical CDF (black points) is shown together with the Poisson-process distribution fit (black line) and the predicted distribution using Kramers' analytical rate (Eq.~\ref{eq:kramers_k0}, cyan dashed line).}
\end{figure}
The rate methods were tested first on the one-dimensional matched-harmonic potential illustrated in Fig.~\ref{fig:results_1d}a given by
\begin{equation}
\begin{aligned}
    U(x)=&\left[\frac{1}{2}\omega^2_0(x-x_0)^2-\frac{\Delta U}{2}\right]\left(1-\Theta(x)\right) \\
    &-\left[\frac{1}{2}\omega^2_1(x-x_1)^2-\frac{\Delta U}{2}\right]\Theta(x)~,
    \label{eq:matchedharmonics}
    \end{aligned}
\end{equation}
where the subscript $0$ corresponds to the well and the subscript $1$ corresponds to the barrier, $\Theta(x)$ is the Heaviside step function, and $\omega_i=\frac{\sqrt{\Delta U}}{x_i}$, which is needed to make the potential continuous. 
Full simulation details for this model are given in Sec.~\ref{sec:methods_1d}.

To estimate the unbiased rate, many Langevin dynamics simulations were performed on the potential from Eq. \ref{eq:matchedharmonics}, starting from the bottom of the well (Fig. \ref{fig:results_1d}a).  The first barrier-crossing time for each simulation was recorded, and the empirical CDF was calculated as described in Sec.~\ref{sec:theory}.
The unbiased rate was extracted by fitting the distribution to the expected Poisson process. 
The 2-sample KS test was performed to assess whether this transition is accurately described by a Poisson process. The $p$-value for the KS statistic is 0.97, demonstrating that the transition times are likely Poisson-distributed data. This yielded a log-rate of $-5.98\pm0.04$ where rates are in units of $\tau^{-1}$, with $\tau$ as the time unit, and the error is the standard deviation obtained from bootstrap analysis\cite{boostrap} (see Sec.~\ref{sec:bootstrap}). The log-rate calculated with Kramers' rate theory using Eq.~\ref{eq:kramers} is $-6.02$, which agrees with the empirical rate within error.

We then performed well-tempered metadynamics (WT-MetaD) simulations (see Sec.~\ref{sec:wtmetad}) for this system to predict the rates using iMetaD, KTR, and EATR using different bias-deposition times $\Delta~dt$ with $dt$ the MD timestep (varying from 10 to 10000 $\tau$, which corresponds to fractions of the mean-first passage time varying from $10^{-6}$ to $10^{-2}$).
In Fig.~\ref{fig:results_1d}, we compare the methods for both the LM and CDF-fit for the situation where $\gamma=1$ is enforced, because (\textit{i}) this allows us to only assess the quality of the different time-dependent rate metrics, and (\textit{ii}) for a 1D potential, all bias should go into promoting the transition as there are no orthogonal degrees of freedom; results obtained from fitting both $k_0$ and $\gamma$ are shown in Fig.~\ref{fig:1D_gamma}. 
We find that the original KTR method does not give a rate consistent with the empirical unbiased rate when $\gamma=1$. On the other hand, we find that both the iMetaD and EATR methods are consistent with the expected values for the rate. 
Moreover, in agreement with Ref.~\citenum{imetadcdf}, we find that fitting the CDF provides more accurate rate estimates than LM at small $\Delta$ for all three methods, with the discrepancy between the two fits negligible for large $\Delta$. The rate estimates for each fitting procedure improve as $\Delta$ increases, which is consistent with the principles of iMetaD. This is also consistent with the results of the KS test. The 2-sample KS test was performed on iMetaD and the 1-sample test was performed for KTR and EATR as explained in Sec.~\ref{sec:ks}. The KS tests failed for the CDF fits at $\Delta~dt=10^1~\tau$ and $10^2~\tau$ and passed for the CDF fits at $\Delta~dt=10^3~\tau$ and $10^4~\tau$. These KS test failures are shown for EATR in Fig.~\ref{fig:results_1d}c as open circles. Given that the rate estimates are more accurate for fitting the CDF (even when the KS test fails), we report results from the CDF-fitting procedure below.

\subsection{Protein G unfolding}
\begin{figure}[h!]
\includegraphics[width=\columnwidth]{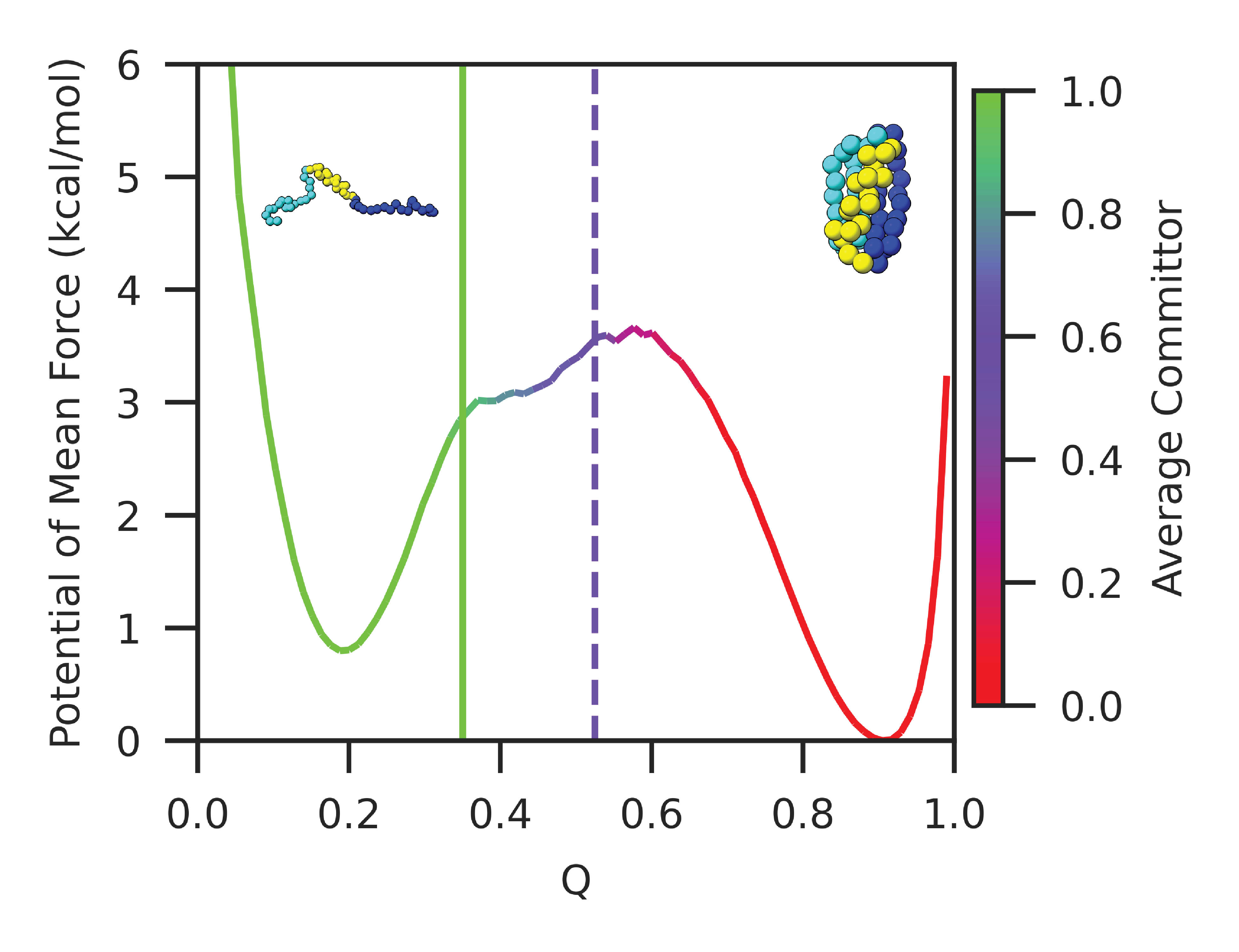}
\caption{ The potential of mean force along the fraction of native contacts $Q$ for the G\=o-like model of the B1 domain of protein G colored according to the average committor function. The value of $Q$ where the average committor is 0.5 is marked with the dashed line, while the critical value for unfolding $Q=0.35$ is marked with the solid line.
\label{fig:qpmfcommit}
}
\end{figure}

\begin{figure*}[ht]
\includegraphics[width=\textwidth]{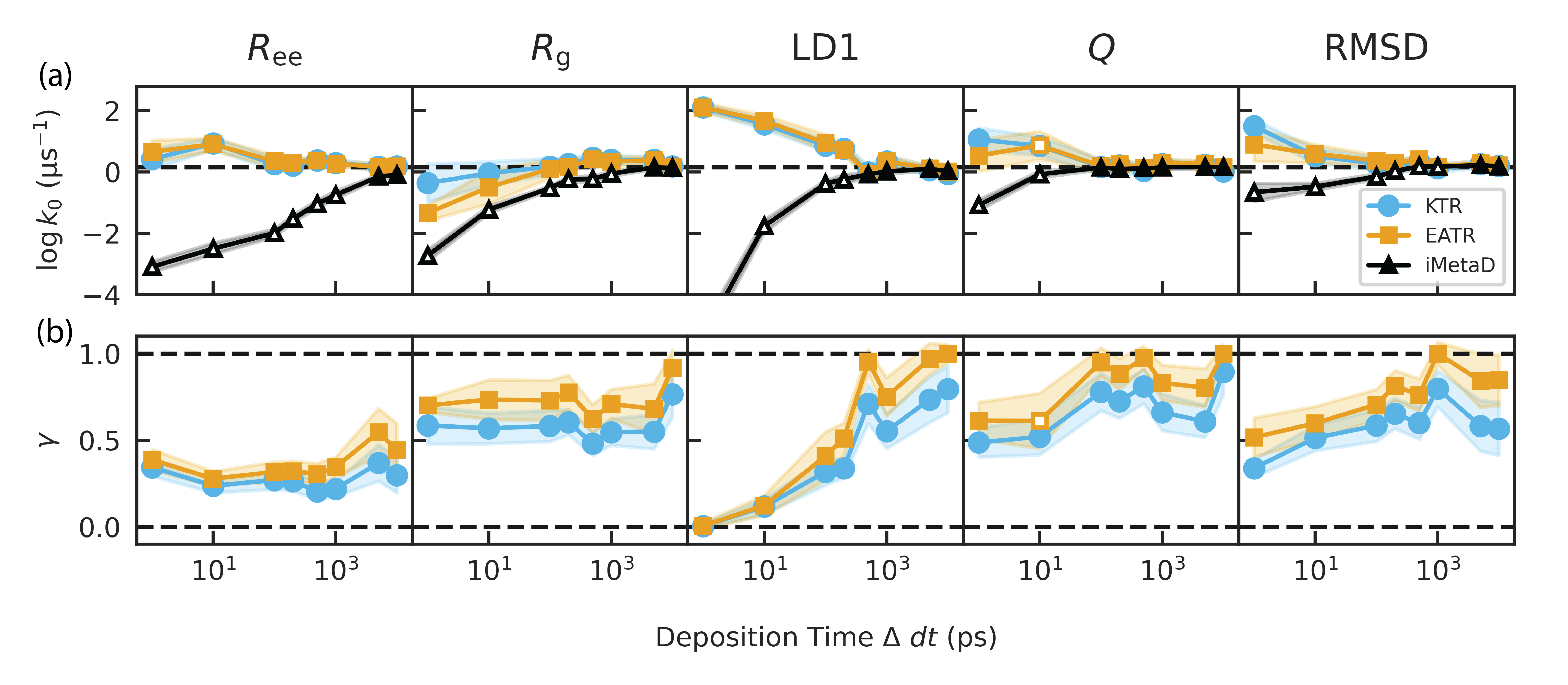}
\caption{\label{fig:ktrfig2} (a) The rates obtained from fitting the CDF for iMetaD (black), KTR (blue) and EATR (orange) for each CV at various deposition times ($\Delta~dt)$. The horizontal dashed line represents the empirical rate obtained from unbiased simulations. Open shapes indicate where the KS test failed. (b) $\gamma$ values obtained for the KTR and EATR methods. Horizontal dashed lines represent the bounds placed on $\gamma$. In both panels, the error bars are computed from a bootstrap analysis as described in Sec.~\ref{sec:bootstrap}.
}
\end{figure*}
We now focus on a more complex system, the unfolding of the B1 domain of protein G using a G\=o-like potential and MD simulations.
This system has the advantage that it is possible to obtain an unbiased estimate of the unfolding rate, while having a rich unfolding landscape complexity, and many possible choices of CVs to characterize the transition \cite{protg1}.    
Below, we will evaluate the quality of several good and bad CVs for predicting rates. 
For this study, we first considered the fraction of native contacts $Q$ and distance between the ends of the protein ($R_\mathrm{ee}$) which were shown to be a good and bad coordinate, respectively, for characterizing the folding of this protein in Ref.~\citenum{protg1}. 
In addition to these two CVs, we will consider the radius of gyration ($R_\mathrm{g}$), the root-mean-squared-deviation from the native state (RMSD), and a recently developed linear-discriminant analysis coordinate maximally separating states as defined through a clustering analysis \cite{posLDA2023} (LD1, see Sec. \ref{sec:cvs}, Fig.~\ref{fig:lda}).

We first performed a 120 $\mathrm{\mu s}$-long unbiased simulation to study the system's behavior. 
For this unbiased trajectory, we computed the potential of mean force (PMF) along each CV by taking the negative log of the histogram of observed CV values (Eq.~\ref{eq:pmf}).
The PMF for $Q$ is shown in Fig.~\ref{fig:qpmfcommit}, and along all CVs in the SI Fig.~\ref{fig:pmfs}, revealing a range of potential profiles and apparent barrier heights. 
Although the PMFs of each CV exhibit two wells, we know that $R_\mathrm{ee}$ is a poor CV for characterizing unfolding because the unfolded ensemble contains configurations with small values of $R_\mathrm{ee}$ comparable to the folded state, resulting in an unusually shaped basin at small values of the CV.

\begin{figure*}[ht]
\includegraphics[width=\textwidth]{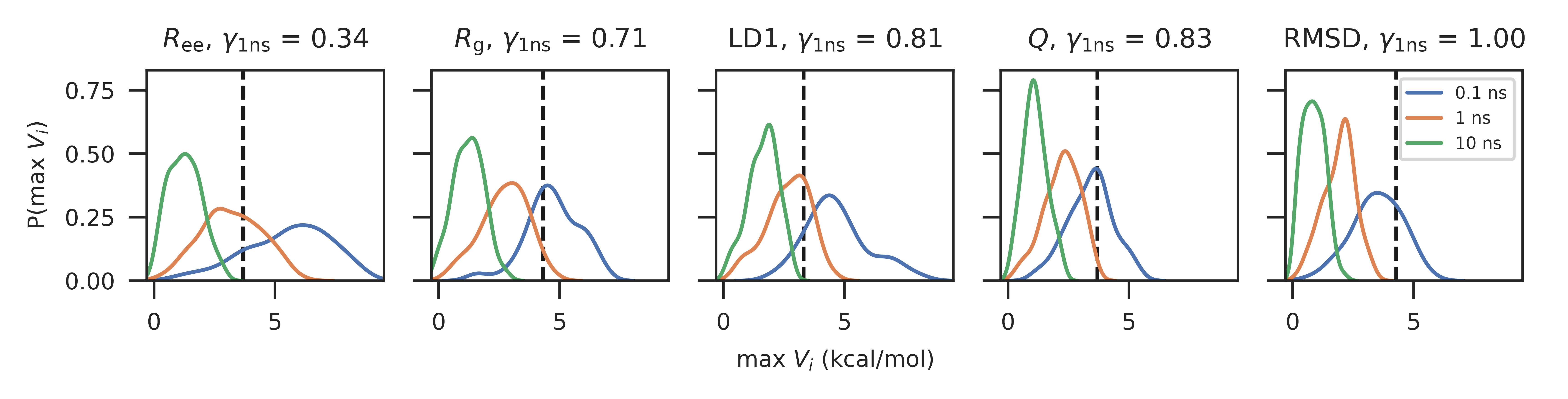}
\caption{\label{fig:bhists} Histograms of the maximum bias deposited for the WT-MetaD simulations for the five CVs and different bias-depositions times. As the biasing efficiency $\gamma$ increases, there is a decrease in the amount of bias needed for transition.}
\end{figure*}

In contrast, we expect $Q$ to be a good CV for unfolding \cite{fracnative}, and so we used $Q$ to define when our system has transitioned out of the folded state by using the average committor in the unbiased simulation \cite{du1998transition,bolhuis2002transition,ma2005automatic}. To do so, we computed for every frame whether the simulation next reached the metastable free-energy minimum on the left (unfolded) before reaching the global minimum on the right (folded), and computed the average by binning these values as a function of the corresponding value of $Q$ in that frame. 
Based on this result, shown in Fig.~\ref{fig:qpmfcommit}, we defined the unfolded region to be where $Q<0.35$ because the average committor to the left of this point is effectively 1.0. To estimate the unbiased rate, we ran 200 unbiased simulations of the protein with randomized initial velocities, and stopped these simulations when $Q$ dropped below $Q=0.35$ using the \texttt{COMMITTOR} function of PLUMED \cite{bonomi2019promoting}. The residence time for the folded state was recorded for each simulation. The unbiased rate was determined to be $1.4 \pm 0.1 \mathrm{\mu s}^{-1}$ using the CDF-fitting procedure previously described. The KS test using a Poisson distribution passed with $p=0.65$, demonstrating a good fit.

We then performed 100 biased simulations for each CV at various bias-deposition times to determine how sensitive each method was to biasing speed. The recovered rates from the methods are shown in Fig.~\ref{fig:ktrfig2}a for each of the biased CVs as a function of the bias-deposition time $\Delta~dt$.
Bias deposition times varied from 1 ps to 10 ns, corresponding to fractions of the mean first passage time ranging from $\sim10^{-6}$ to $\sim10^{-2}$ as in the case of the simple potential in the previous section. 
As expected, longer hill deposition times are observed to generally increase the accuracy of all rate calculations. However, for intermediate to fast-deposition times, KTR and EATR extract unbiased rates closer to the true rate than does iMetaD, especially for the three CVs shown on the left of Fig.~\ref{fig:ktrfig2}a.
We also performed a similar study using untempered MetaD and find that, similarly, all methods work well, with KTR and EATR slightly out-performing iMetaD in the fast biasing regime (SI Fig.~\ref{fig:untemp}).

The KTR and EATR methods also give a measure of the CV biasing efficiency $\gamma$, which is shown in Fig.~\ref{fig:ktrfig2}b. We find that $Q$ and RMSD typically give higher values of $\gamma$ than the end-to-end distance ($R_\mathrm{ee}$) and the radius of gyration ($R_\mathrm{g}$). This coincides with the physical intuition of protein unfolding CVs, where the number of contacts and the similarity to the folded structure should be most relevant. This also agrees with Ref.~\citenum{fracnative}, which found that $Q$ is a good CV for this system.
The CV obtained from linear discriminant analysis (LD1, Ref.~\citenum{posLDA2023}) appears to have a large value of $\gamma$ for slow biasing and a small value of $\gamma$ for fast biasing. A similar trend appears for all CVs tested, but this is most prominent in LD1, and we are still investigating the reason $\gamma$ for LD1 is so much more sensitive than the other CVs here, while still serving as a very good CV for distinguishing folded and unfolded states (as proposed in our previous study \cite{posLDA2023}).
We note that the discrepancy between iMetaD and the true rate is most pronounced when KTR or EATR predict lower values of $\gamma$.

Our intuitive expectation is that a bad CV would require significant amounts of extra bias to be deposited before the system can overcome the apparent barrier in the FES for that CV, necessitating a low value of $\gamma$ to compensate in our rate calculation. 
To check this, we computed the histogram of the maximum bias across the different simulations at different deposition rates. 
CVs with high $\gamma$ and slow deposition have maximum bias that do not exceed the apparent barrier, while fast biasing and poor CVs require substantial extra energy to be injected into the system to effect transitions. 
Here, we use $\max V_i$ because it is a direct ingredient in the KTR method. 
We note that another interesting quantity to compute here would be the amount of non-equilibrium work performed by the MetaD bias, which has been recently exploited in another estimator of rates from time-dependent biased simulations \cite{kawarate}.

For all these CVs, the KTR and EATR methods performed comparably well and consistently performed better than the iMetaD method. Interestingly, the KS test seems not to be as sensitive for KTR and EATR as it was in iMetaD. 
Fitting the CDF for iMetaD results in failed KS tests even where the error in the rate is small, but KTR never failed the KS test for these CVs and EATR only failed for one condition.
This may be an effect of introducing $\gamma$ as an additional fitting parameter, so it is possible to get fits closer to the empirical CDF with worse rate estimates. 
In Fig.~\ref{fig:gamma_ranges}, we show that the introduction of  $\gamma$ allows us to make good fits to the CDF; indeed, many pairs of $(k_0, \gamma)$ can be used to fit these data; however, we note that the resulting predicted rates are still quite close to the most confident rate prediction, so this small amount of flexibility is not a problem here in practice.

\subsubsection{2D MetaD}
\begin{figure}[ht]
\includegraphics[width=\columnwidth]{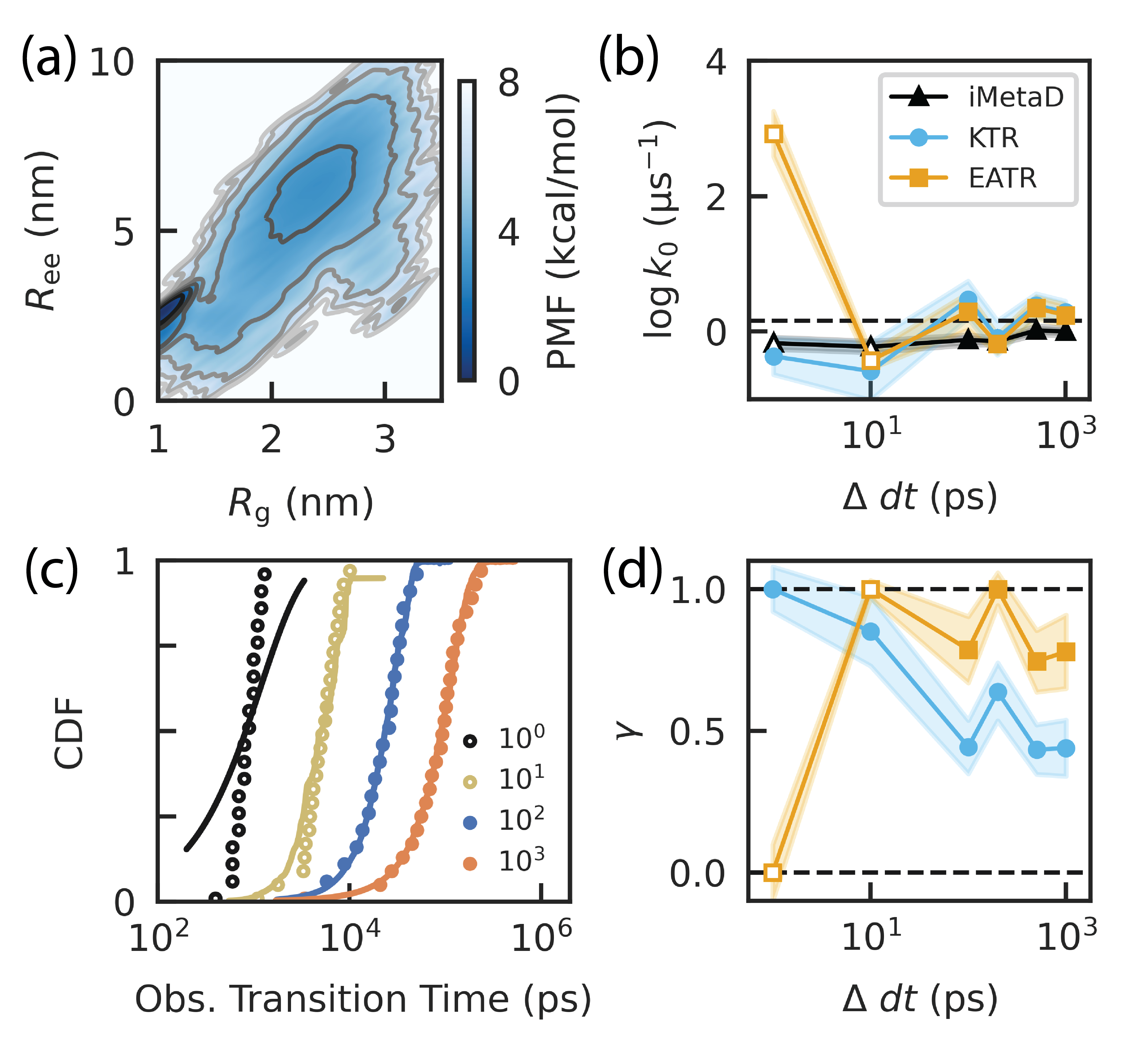}
\caption{\label{fig:results_2d} (a) The potential of mean force projected onto the 2D $R_\mathrm{g}$/$R_\mathrm{ee}$ space. (b) The rates obtained from each method at various deposition times. The horizontal dashed line represents the true unbiased rate for this system. (c) The empirical and EATR-fit CDFs for each deposition time. The deposition times in the legend are given in ps. (d) The $\gamma$ values obtained for KTR and EATR at various deposition times. In panels b, c, and d, open shapes indicate where the KS test failed.
We performed some simulations with even slower biasing with the result that not every simulation transitioned; while the resulting rate estimate was unaffected, the estimated value of $\gamma$ dropped significantly. The effects on the rate and $\gamma$ are shown in SI Fig.~\ref{fig:incomplete}.
}
\end{figure}

MD studies involving large biomolecules or molecular assemblies will typically have many slow degrees of freedom characterizing transitions between important states, and hence we expect the need to use multiple CVs to bias the system in order to promote transitions in a reasonably short amount of simulation time.
We wanted to assess whether KTR and EATR still work for this case, despite the fact that the role of $\gamma$ in characterizing CV quality is less direct. 
To do so, we performed MetaD simulations while simultaneously biasing the end-to-end distance $R_\mathrm{ee}$ and the radius of gyration $R_\mathrm{g}$. For these two CVs, a combined PMF from the long unbiased trajectory is shown in Fig.~\ref{fig:results_2d}a.  
The rates obtained from each rate method are shown in Fig.~\ref{fig:results_2d}b, all methods recovered the rate equally well apart from EATR for the fastest bias-deposition time, where the KS test failed.
The corresponding CDFs and EATR fits are shown in Fig.~\ref{fig:results_2d}c, confirming that the method is able to predict the distribution of transition times.
For intermediate $\Delta~dt\ge10^2~\mathrm{ps}$, the value for $\gamma$ improved in this case over only biasing $R_\mathrm{ee}$ and improved slightly over only biasing $R_\mathrm{g}$. This suggests that biasing multiple CVs simultaneously increases the efficiency of biasing without affecting the rate estimation. This might also be the reason for why iMetaD preforms well in this case, while it failed for the same bias-deposition times in the 1D biasing cases (see Fig. \ref{fig:ktrfig2}).

\section{Conclusions}

In this work, we have developed a general rate theory for time-dependent biased simulations that encompasses several of the existing methods by using the bias scaling factor $f(t)$ (from Eq. \ref{eq:gen_rate_estimator}) with different analytical formulations. In practice, LM and CDF-fitting are two different manners for estimating the unbiased rate from an ensemble of simulations launched from a single state.
Although we demonstrate in this work that the previously proposed KTR approach works robustly for a realistic problem, it does not predict as accurate results as iMetaD for the case of an ideal 1D CV. Therefore, we proposed the EATR formulation, which exactly coincides with iMetaD in the ideal CV case but which gives good rate estimates for bad coordinates more robustly than does iMetaD, with the additional benefit of reporting on the efficiency of that parameter through the parameter $\gamma$.  

We have validated the methods over the more complex landscape of G-protein unfolding where we could still have ground truth. We have found that both the KTR and EATR methods offer accurate rate measurements from biased simulations. The accuracy of the rates determined from these methods are surprisingly insensitive to biasing rate and CV quality, even for frequent-biasing regimes where the average maximum bias likely exceeds the true free energy barrier.
We find that the $\gamma$ computed from both KTR and EATR report CV efficiencies $\gamma$ that correlate with our qualitative intuition of what is a good or bad biasing coordinates. 
Overall, we find that the CDF fit is better than LM to obtain the rate, however, many solution pairs of $\{\gamma,k_0\}$ could pass the KS test, so getting a good fit is not sufficient to guarantee that the optimal unbiased rate was obtained; however, in practice the predicted rates were all very close to the highest confidence prediction (as shown in SI Fig.~\ref{fig:gamma_ranges}). 
We find that for 2D biasing landscapes all methods perform reasonably well, which is an indication that biasing many directions might be helpful for barrier-crossing enhancement. In the future, we hope to test the methods on more complex systems using multiple biasing dimensions and, if necessary, extend the theory to multiple dimensions as was done for force-spectroscopy in Refs.~\citenum{cossio2015artifacts} and \citenum{cossio2018transition}. 

Despite this success, there are several aspects of the EATR method that can still be improved through future work. 
For example, we could take into account the effect of the bias on the pre-exponential factor. 
Although the method works well in the case of our coarse-grained model of protein G for very fast biasing, we have not solved the general problem of how to compute rates in the over-biasing regime where $\gamma$ times the bias could still be larger than the true barrier, as for example for LDA in Fig. \ref{fig:bhists} at fast deposition times, which could lead to the overestimated rate and small $\gamma$ (Fig. \ref{fig:ktrfig2}). 
Addressing this issue will be crucial for systematically using the EATR method for large systems with many slow degrees of freedom.
Going forward, we would like to determine whether it is possible to find a theoretical interpretation for $\gamma$ in multiple dimensions, e.g. whether it can be derived considering projection operator approaches, and investigate whether it is rigorously connected to non-equilibrium estimators of the effect of a time-varying bias on the rates \cite{kawarate}. 
Finally, we note that in this work we have concentrated on MetaD as a way of applying bias, but given our new generalized rate estimator, it would be interesting to compare whether there are other more efficient CV-based biasing protocols that give similarly accurate results.

\section{Computational Methods}

\subsection{Overdamped Langevin dynamics on a 1D potential}
\label{sec:methods_1d}

We ran overdamped Langevin dynamics over the potential given in Eq.~\ref{eq:matchedharmonics} using $x_0=-3$, $x_1=3$, and $\Delta U=8~\mathrm{k_BT}$. We used an integration time step of $dt=$0.01~$\tau$ where $\tau$ is the time unit. The friction parameter used was $0.02~\tau$, which corresponds to the friction coefficient $\zeta=50~\tau^{-1}$ and the diffusion coefficient $D=0.02~\tau \mathrm{k_BT}$. All simulations were started from $x=-3$. 
Because the diffusion coefficient and potential are known, we can derive the standard Kramer's expression in the Smoluchowski limit \cite{hanggi1990reaction} from Eq.~ \ref{eq:kramers_k0}  
\begin{equation}
    k_0=\frac{\omega_0\omega_1}{2\pi \zeta}e^{-\beta \Delta U}~,
    \label{eq:kramers_k0}
\end{equation}
and use that to determine the unbiased rate.
For this specific system, the theoretical rate is $9.49\times10^{-7}~\tau^{-1}$.
These Langevin dynamics simulations were performed using the PESMD tool in PLUMED \cite{bonomi2019promoting}. Some of these simulations were biased using WT-MetaD with a starting hill height of 1 $\mathrm{k_BT}$, a $\sigma$ of 0.5, and a biasfactor of 2.0. MetaD simulations were performed for four bias-deposition times ($\Delta~dt$): $10^1~\tau$, $10^2~\tau$, $10^3~\tau$, and $10^4~\tau$.

\subsection{Well-tempered metadynamics}
\label{sec:wtmetad}

In WT-MetaD simulations, a history-dependent biasing potential $V(\xi,t)$ is generated at a position $\xi$ in CV-space.
$V(\xi,t)$ is formed as a sum of Gaussians with width $\sigma$ (which can differ for each CV) and height $h$ deposited every $\Delta$ steps. 
For a one-dimensional CV, this can be written as,
\begin{equation}
    V(\xi,t)= \sum_{j=1}^{N_\mathrm{hills}} h\;e^{-\frac{V(\xi,t_j)}{k_B \Delta T}}e^\frac{-\left(\xi(t_j)-\xi\right)^2}{2\sigma^2}~,
    \label{eq:Vmetad}
\end{equation}
where $t_j=j\Delta~dt$ are the times where hills were deposited prior to time $t$, and $N_\mathrm{hills}=\lfloor t/(\Delta~dt) \rfloor$. Here $\Delta T$ is a tempering factor which causes the heights to decrease proportionally to how much bias is already applied at that point, and is specified in PLUMED by setting a biasfactor of the form $(T+\Delta T)/T$, where $T$ is the simulation temperature.
In the original untempered MetaD, the hills are of constant height, i.e. $\Delta T\rightarrow\infty$.
In iMetaD, the pace $\Delta~dt$ would be taken to be large, such that the frequency of deposition $(\Delta~dt)^{-1}$ becomes small. For notation simplicity, we have omitted the explicit dependence on $\xi$ from Eq.~\ref{eq:Vmetad} in all equations in the Theory.

\subsection{G\={o}-like model of protein G}

A G\=o-like coarse grained model of the B1 domain of protein G was prepared to assess the accuracy of the rate extraction methods, starting from PDB ID 1PGB. This system was selected because it was previously used as a paradigmatic example of a two state folder with known good and bad reaction coordinates \cite{kb2002,protg1,protg2}.
In a G\={o}-like model, each residue is modeled as a bead at the position of the $\alpha$-carbon. The force field for this model treats pseudo-bonds and angles harmonically, and pseudo-dihedrals using a Fourier series. Non-covalent interactions, as in Ref.~\citenum{kb2003}, depend on whether the residues are in contact in the native structure, which is determined by whether the side chains of two residues contain heavy atoms within 4.5$\textup{~\AA}$ of each other. The force field parameters for the model in Refs.~\citenum{protg1,protg2} were provided by the authors.
Our implementation of the potential in LAMMPS \cite{thompson2022lammps} and input files for all simulations are provided in the GitHub for this article (see Sec.~\ref{sec:data}).

\subsubsection{Molecular dynamics simulations}

The MD simulations of the G\=o-like model were performed using the LAMMPS software package \cite{thompson2022lammps}. The software was updated partway through the project and the version used for each set of simulations is shown in Table~\ref{tab:software_versions}. All simulations used a time step of $dt=$10 fs, and the temperature was held constant at 312 K using the Nos\'{e}-Hoover chain thermostat \cite{martyna1992nose} with a damping factor of 1 ps and a chain length of 3. All simulations started from the folded structure.

For the unbiased simulations, we ran 200 replicates and ended the simulations when the protein model unfolded, which was defined to be when the CV $Q$ decreased past 0.35 as described above. The empirical CDF for the transition times to the unfolded state were fit to Eq.~\ref{eq:poisson_survival} as explained in the Theory section to obtain the observed unbiased rate for this system.

\subsubsection{Collective Variables}
\label{sec:cvs}

A variety of CVs were analyzed for the G\=o-like model, which were used for the biased simulations and during the rate analysis. The first of these is the fraction of native contacts ($Q$), which captures the degree to which the protein is folded. This CV is the fraction of the contacts present in the native structure which are still present, and was defined as in Ref.~\citenum{protg2}. The end-to-end distance ($R_\mathrm{ee}$) was also used, as it was previously determined to be a poor coordinate\cite{protg1}. The RMSD of the protein with respect to the native structure and the radius of gyration ($R_\mathrm{g}$) were included to compare with the previously used CVs for this system.

To define the LD1 coordinate, first we performed a cluster scan on the unbiased trajectory using the shapeGMM clustering algorithm \cite{gmm} with 50,000 frames for training, 3 training sets and 15 attempts each, for cluster sizes ($K$) = 2,$\dots$,6. The training curve with cross validation from the scan is shown in Supplementary Fig.~\ref{fig:lda}. We used the positions of the beads as input features for shapeGMM. We did a 5 state shapeGMM fit on the entire trajectory ($\sim$ 1.2M frames) with 15 attempts to identify the distinct clusters. We then performed an iterative global alignment of the trajectory to the global mean and covariance. Multi-state Linear Discriminant Analysis (LDA) was performed on the globally aligned trajectory with frames from all 5 clusters. Only the first coordinate (LD1) out of four resulting LD coordinates has been used in this study.\cite{posLDA2023} 
In Supplementary Fig.~\ref{fig:lda}b,c we show that this coordinate completely separates the folded and unfolded states, with the other states appearing as intermediates. 

\subsubsection{Biased simulations}

The collective variables and biasing for protein G were handled using PLUMED \cite{bonomi2019promoting}. The version of PLUMED used for each set of simulations is shown in Table~\ref{tab:software_versions}. As is the case for the 1D potential, WT-MetaD was used to bias the simulations. A set of untempered MetaD simulations were also performed, the results of which are provided in the Supplementary Information. The  parameters used for the WT- and untempered MetaD simulations are given in Supplementary Tab.~\ref{tab:metad_params}. The values of $\sigma$ were chosen for WT-MetaD according to the standard deviation of the biased CV in the folded state, and for untempered MetaD $\sigma$ was chosen to be less than that used in WT-MetaD. 
We performed simulations at eight different bias deposition times ($\Delta~dt$): 1 ps, 10 ps, 100 ps, 200 ps, 500 ps, 1 ns, 5 ns, and 10 ns. 100 simulations were performed for each $\Delta~dt$. The simulations were halted when the protein was determined to have unfolded, or when either wall-clock time reached 48 hours or a total simulation time of 10 $\mu$s was reached.

\subsubsection{Potential of mean force and committor analysis}
A long simulation of protein G was performed and the potentials of mean force (PMFs) along various CVs were determined from the unbiased simulation data using
\begin{equation}
    A(\xi)=-k_BT\hspace{1pt}\log\,P(\xi)~,
    \label{eq:pmf}
\end{equation}
where $\xi$ is the CV along which the potential of mean force is computed and $P(\xi)$ is the probability density of $\xi$ obtained by computing a normalized histogram.

Committor analysis\cite{du1998transition,bolhuis2002transition,ma2005automatic} along $Q$ was done on this long simulation by assigning either 0 or 1 to each frame of the trajectory depending on whether the system visits the folded or unfolded state next, then taking the average for all frames associated with each value of $Q$. In order to prevent incorrect assignments to either state, for the committor analysis the system was considered to be in the unfolded state when $Q<0.25$ and to be in the unfolded state when $Q>0.85$. From this, $Q=0.35$ was decided to be the critical value for unfolding, as illustrated in Fig.~\ref{fig:qpmfcommit}. This was chosen to be less than the transition state to prevent counting cases where the system enters the transition region, but fails to unfold.

\subsection{Bootstrap analysis}
\label{sec:bootstrap}
Errors were obtained from bootstrap analysis\cite{boostrap}. For this analysis, a new set of transition times was constructed by choosing random simulations from the original set with replacement. Once the new set had the same size as the original set, the rate calculation was performed on the new set. This was repeated 100 times and the standard deviation of the log of the rate and $\gamma$ across these new sets is reported.

\subsection{Kolmogorov-Smirnov test}
\label{sec:ks}

The KS test was performed to assess whether the transition distribution is accurately described by the expected theoretical distribution. For the case of unbiased or iMetaD, it is a Poissonian distribution. The 2-sample KS test was used for the unbiased and iMetaD analyses. This version of the test determines the maximum deviation of the observed CDF from two samples and gives a $p$-value which, when sufficiently low, allows us to conclude that the samples most likely did not come from the same underlying distribution. We consider the empirical and theoretical distributions to coincide if $p > 0.05$. The 1-sample KS test was used for the KTR and EATR analyses, as generating large random samples from their distributions took a significant amount of time. This version of the test determines the maximum deviation of the observed CDF for one sample from a theoretical CDF, and gave the same results as the 2-sample test in all the cases that were checked.

\section*{Data availability}
\label{sec:data}
Inputs for simulations, LAMMPS code for the G\={o}-like model, code for analysis, data for each system consisting of the value of the CVs and bias versus time, a script for clustering and generating the LDA coordinate, and code for generating figures are available at \url{https://github.com/hocky-research-group/EATR-paper-2024}.

\section*{Acknowledgements}

We thank K. Palacio-Rodriguez, F. Pietrucci, L.S. Stelzl, and A. Szabo for discussing this research with us and offering ideas for future directions. We are grateful to R.B. Best for providing us with his non-covalent force field parameters for the protein G G\=o-like model.

NM, SS, and GMH were supported by the National Institutes of Health under award number R35GM138312. SS was also partially supported by a graduate fellowship from the Simons Center for Computational Physical Chemistry (SCCPC) at NYU (SF Grant No. 839534). This work was supported in part through the NYU IT High Performance Computing resources, services, and staff expertise, and simulations were partially executed on resources supported by the SCCPC at NYU.
PC was supported by the Flatiron Institute, a division of the Simons Foundation. NM thanks the Center for Computational Mathematics and GMH the Center for Computational Biophysics at the Flatiron Institute for their hospitality while a portion of this research was carried out.

\bibliography{biblio}

\onecolumn
\begin{center}
\LARGE \textbf{SUPPORTING INFORMATION} 
\end{center}
\setcounter{section}{0}
\renewcommand{\thesection}{S\arabic{section}}
\setcounter{figure}{0}
\renewcommand{\thefigure}{S\arabic{figure}}
\renewcommand{\thetable}{S\arabic{table}}

\section{iMetaD from the general time-dependent rate theory}

Let simulation $i$ have a history-dependent bias $V_i(t)$ which scales the rate for that simulation as
\begin{equation}
    f_i(t)=e^{\beta V_i(t)}~,
\end{equation}
where the ``rate for a simulation" is the inverse of the mean observed transition time we would expect for a set of simulations experiencing the same bias potential. Substituting this into Eqs.~\ref{tdr_poisson_survival} and \ref{eq:gen_rate_estimator} yields
\begin{equation}
    S_i(t_i) = e^{-k_0\int_0^{t_i} e^{\beta V_i(t')}\,dt'}~,
\end{equation}
with
\begin{equation}
    k^*_0=\frac{M}{\sum_{i=1}^N\int_0^{t_i}e^{\beta V_i(t')}\,dt'}~.
\end{equation}
If all simulations transition and we define the acceleration factor $\alpha_i$ to be
\begin{equation}
    \alpha_i = \frac{1}{t_i} \int_0^{t_i} e^{\beta V_i(t')}\,dt'~,
\end{equation}
we can reinterpret the scaling of the rate as a scaling of time by $\alpha_i$
\begin{equation}
    S(t_i) = e^{-k_0 \alpha_i t_i} = e^{-k_0 t_i^{rescaled}}~,
\end{equation}
with
\begin{equation}
    k^*_0=\frac{N}{\sum_{i=1}^N\alpha_it_i}=\frac{1}{\left< t_i^{rescaled} \right>}~,
\end{equation}
which are the survival function and rate estimator for iMetaD, respectively.

\section{Results for a matched harmonic potential where $\gamma$ is fit rather than being restricted to $\gamma=1$}

\begin{figure*}[h!]
\vspace{-2em}
\includegraphics[width=\textwidth]{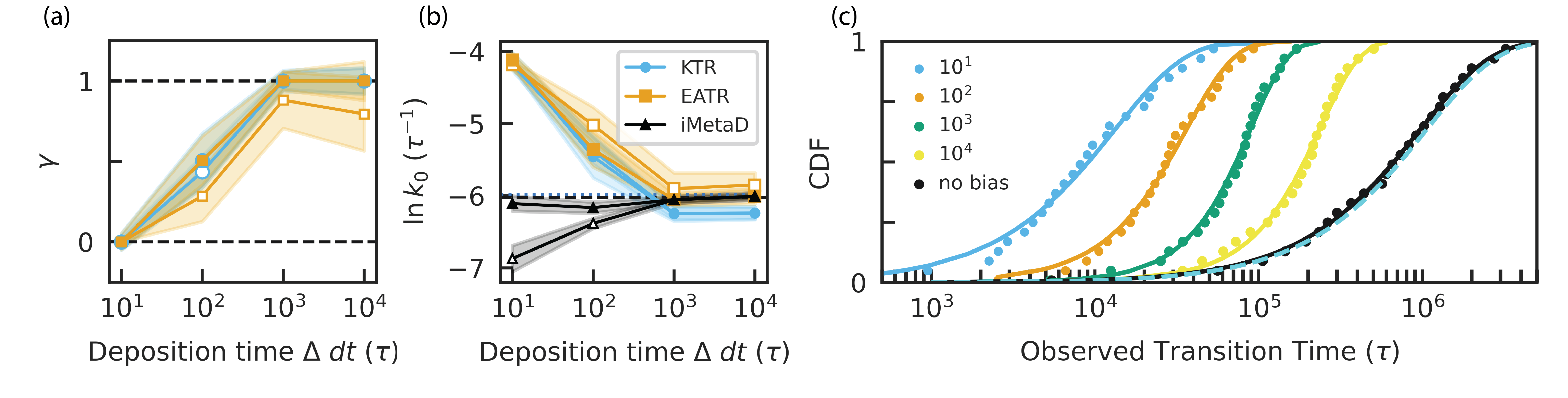}
\vspace{-2em}
\caption{\label{fig:1D_gamma} (a) $\gamma$ obtained from the KTR and EATR methods, where the maximum likelihood estimates are represented by open symbols and the least-squares CDF fits are represented by filled symbols. (b) The rates obtained from the iMetaD, KTR, and EATR methods. iMetaD provides better rate estimates in this case because this is a perfect CV, which iMetaD assumes while KTR and EATR do not. (c) The best-fit EATR CDFs for each value of deposition time, $\Delta~dt$.}
\end{figure*}

\clearpage
\section{ShapeGMM and Position-LDA clustering analysis}
\begin{figure*}[h!]
\vspace{-2em}
    \includegraphics[height=2.5in, width=7in]{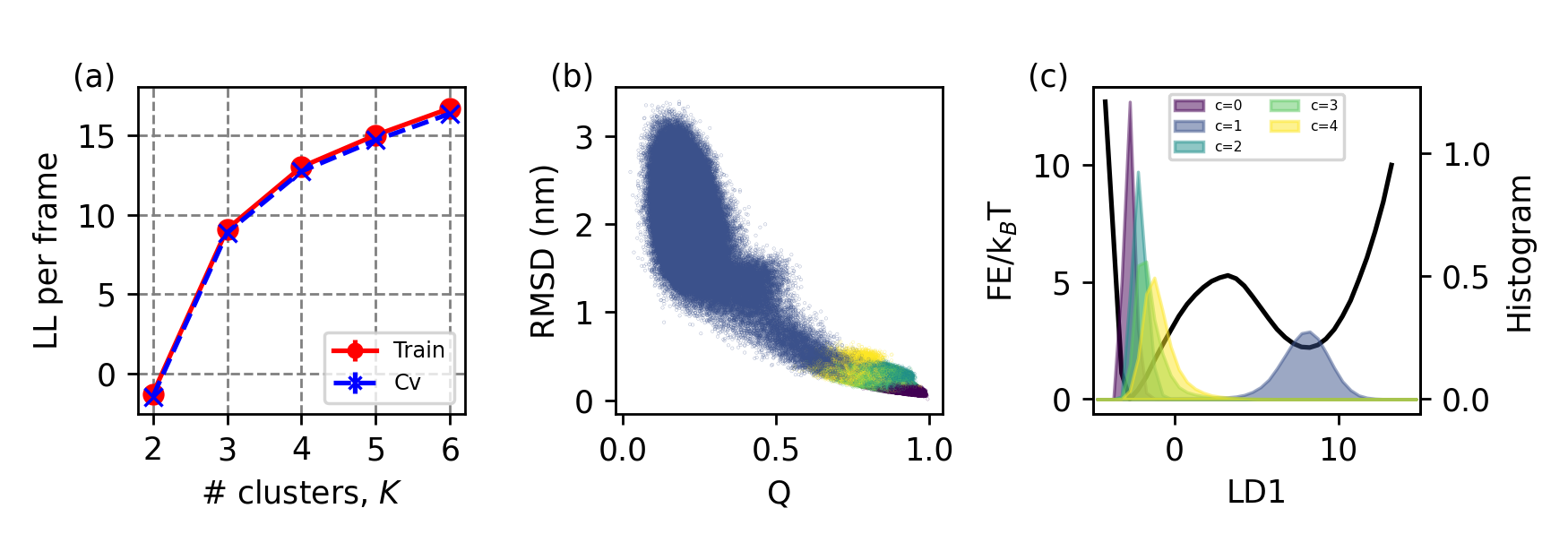}
    \vspace{-2em}
    \caption{(a) Log-likelihoods of data coming from a shapeGMM model trained on $K$ clusters as described in Sec.~\ref{sec:cvs}. 
    The agreement between training and cross-validation prediction suggests that the data is not over-fit for any of these numbers of clusters.
    (b) 2D scatter plot of sampled conformations colored according to their cluster assignments from ShapeGMM using a $K=5$ model. Based on the position of each cluster in $(Q,$RMSD$)$ space, we assign the 0 to be the folded state and 1 the unfolded state (colors defined in c). (c) 1D PMF profile (solid black) along the LD1 coordinate. Histograms of LD1 corresponding to separate clusters, normalized individually and colored according to cluster id are shown in transparent colors.}
    \label{fig:lda}
\end{figure*}

\section{PMFs calculated for G\={o}-like model of Protein G from an unbiased trajectory}
\begin{figure*}[h!]
\vspace{-2em}
\includegraphics[width=\columnwidth]{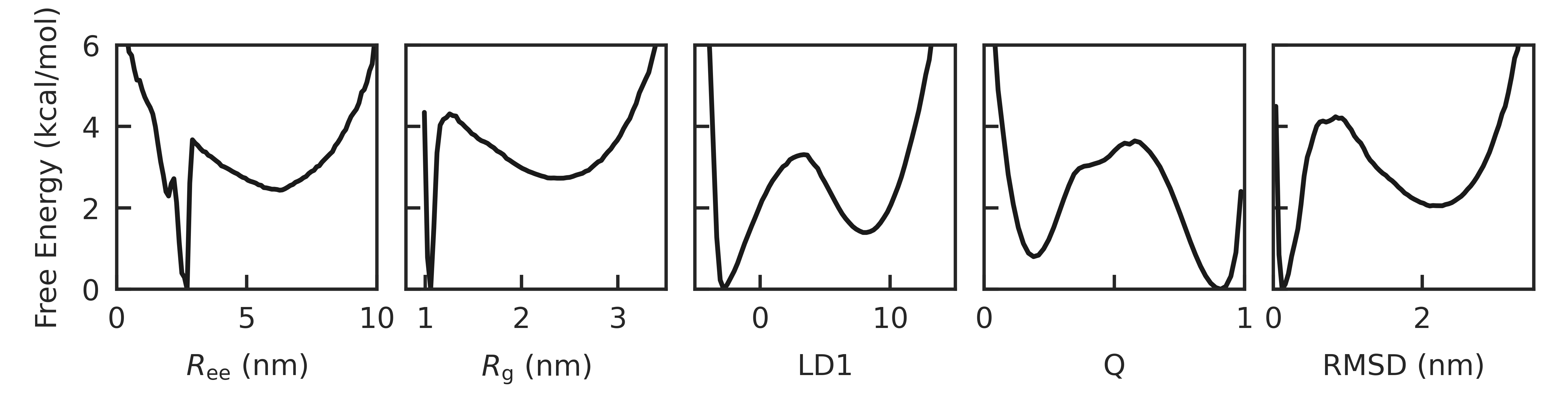}
\vspace{-2em}
\caption{\label{fig:pmfs} PMFs of the G\=o-like model of Protein G computed along each of the five CVs presented in the Computational Methods section.}
\end{figure*}

\clearpage
\section{Rate estimates using untempered MetaD}
\begin{figure*}[h!]
\vspace{-2em}
\includegraphics[width=\textwidth]{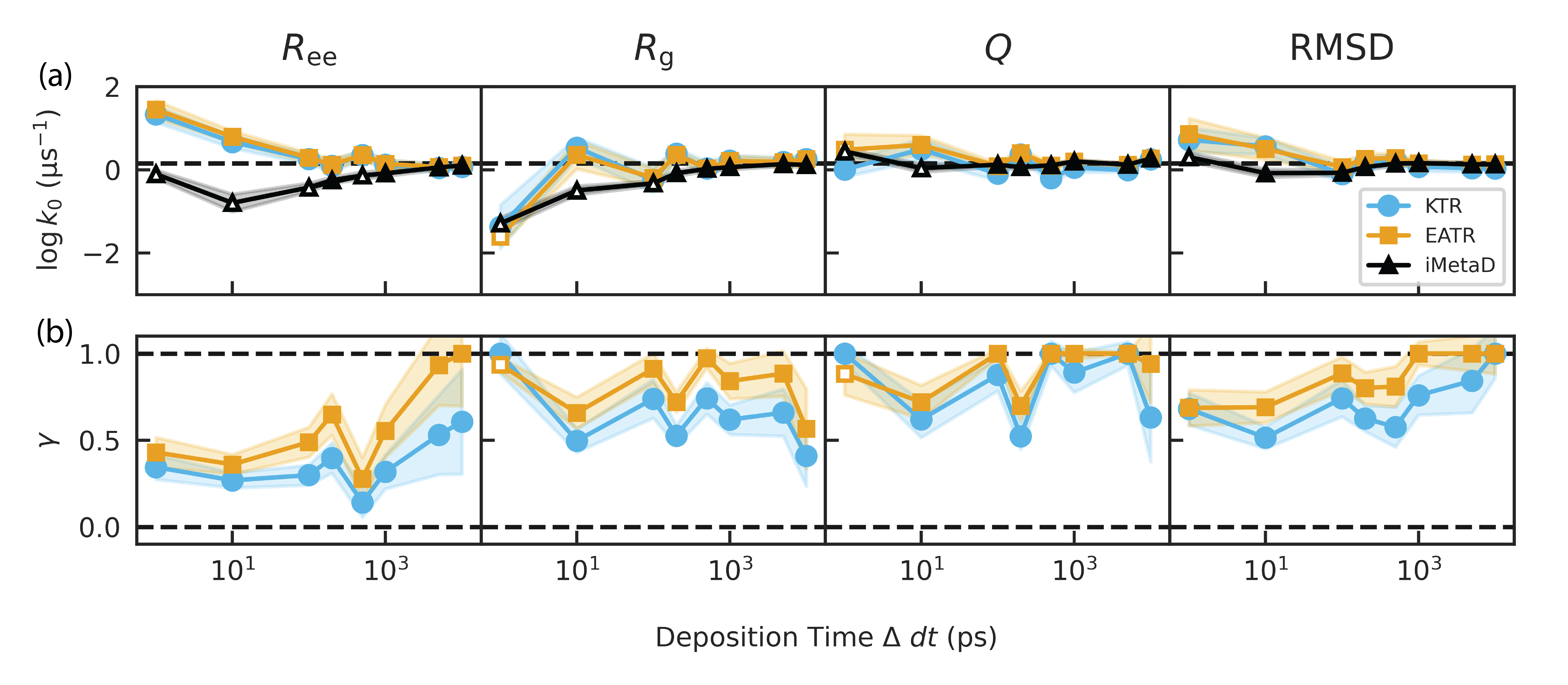}
\vspace{-2em}
\caption{\label{fig:untemp} (a) The rates obtained from the four CVs biased using untempered MetaD using the iMetaD, KTR, and EATR methods. These methods provide accurate results in this biasing scheme. (b) The values of $\gamma$ obtained from the KTR and EATR methods for each of the CVs.}
\vspace{-2em}
\end{figure*}

\section{Multiple $k_0$ and $\gamma$ pairs pass the KS test}
\begin{figure*}[h!]
\vspace{-2em}
\includegraphics[width=\textwidth]{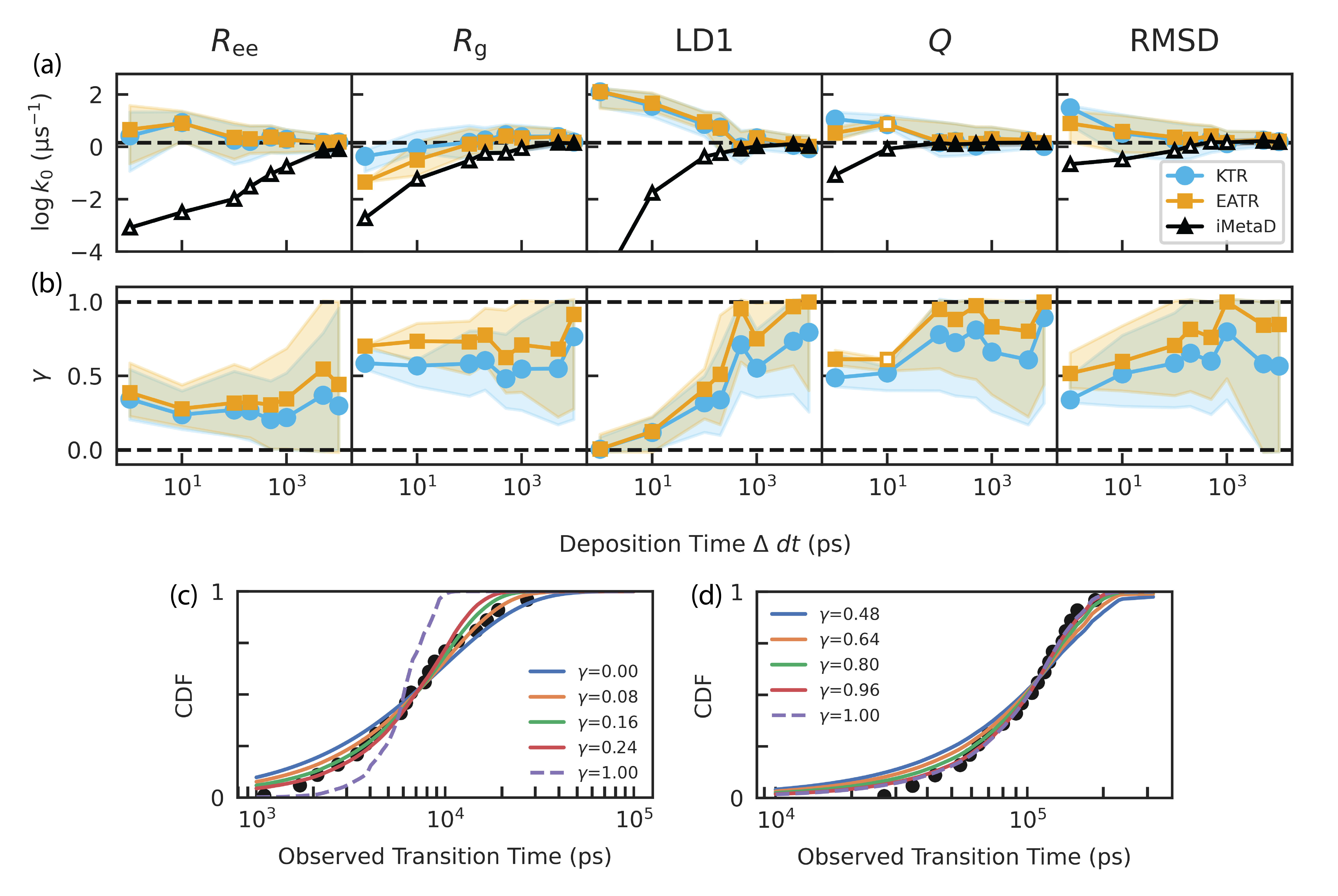}
\vspace{-2em}
\caption{ Same data as Fig.~\ref{fig:ktrfig2}. (a) Shaded regions represent the ranges of rates for each of the five CVs from the iMetaD, KTR, and EATR methods which pass the KS test for some value of $\gamma$. (b) Shaded regions represent the ranges of $\gamma$ values from the KTR and EATR methods which pass the KS test for some value of the rate. (c) Several EATR CDFs which pass the KS test for the LD1 CV at $\Delta~dt=10~\mathrm{ps}$. (d) Several EATR CDFs which pass the KS test for the RMSD CV at $\Delta~dt=1~\mathrm{ns}$. }
\label{fig:gamma_ranges}
\vspace{-2em}
\end{figure*}

\clearpage
\section{Effect of un-transitioned simulations on $k_0$ and $\gamma$}

\begin{figure*}[h!]
\includegraphics[width=\textwidth]{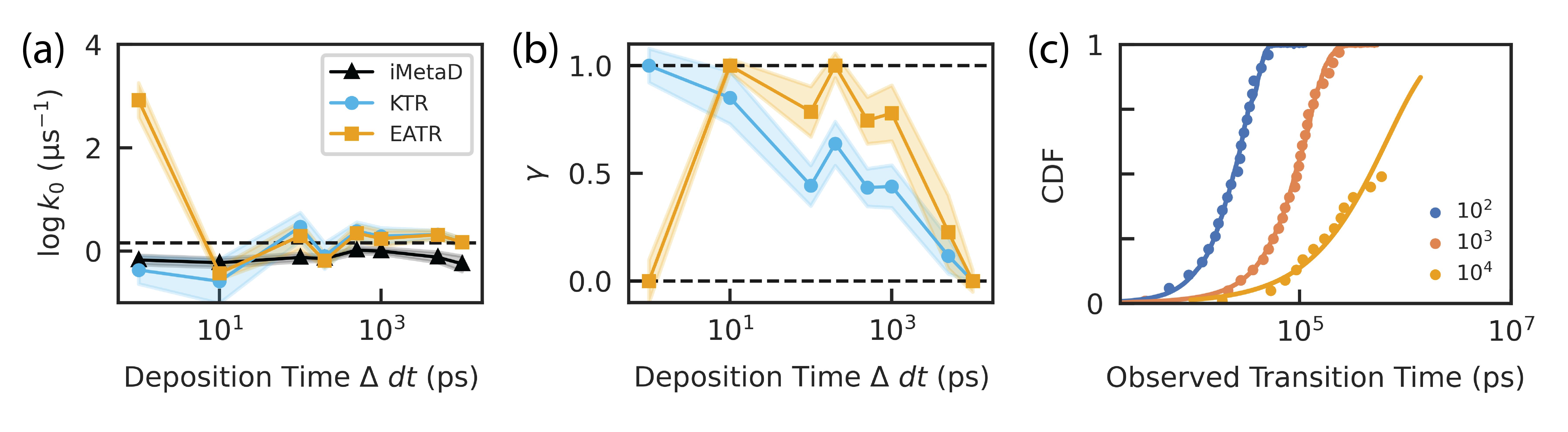}
\caption{ Same as Fig.~\ref{fig:results_2d}, with additional slower deposition times where not all simulations transitioned. (a) The rates obtained from fitting the CDF for iMetaD, KTR, and EATR for the 2D $R_\mathrm{ee}$/$R_\mathrm{g}$ CV. The rate estimates for $\Delta~dt=5~\mathrm{ns}$ and $10~\mathrm{ns}$ are as accurate as the previous points, although only 66\% and 50\% of the simulations transitioned, respectively. (b) The values of $\gamma$ obtained from KTR and EATR. The $\gamma$ estimates for $\Delta~dt=5~\mathrm{ns}$ and $10~\mathrm{ns}$ are significantly lower than the previous points. (c) The empirical CDF and the best fit EATR CDF for $\Delta~dt=100~\mathrm{ps}$, $1~\mathrm{ns}$, and $10~\mathrm{ns}$. The KS test is not reported for incomplete simulation sets due to how the KS test is implemented in SciPy. }
\vspace{-2em}
\label{fig:incomplete}
\end{figure*}

\section{Simulation software and parameters}
Simulations were initially performed with the LAMMPS version from 5 May 2020 and PLUMED 2.6.6. We switched to a newer version of LAMMPS and PLUMED to perform simulations with the LDA coordinate, and due to an error in how the older version of PLUMED was computing $Q$, which we corrected in PLUMED 2.8 \footnote{\url{https://github.com/plumed/plumed2/pull/951}}.
The exact versions used for all of our simulations are given in Tab.~\ref{tab:software_versions}.

\begin{table}
\begin{tabular}{|c|c|c|}
    \hline
    Simulations & LAMMPS & PLUMED  \\
    \hline
    Unbiased 1D & --- & v2.8.1 \\
    \hline
    Biased 1D & --- & v2.8.1 \\
    \hline
    Unbiased GB1 & 5 May 2020 & --- \\
    \hline
    $R_\mathrm{ee}$ & 5 May 2020 & v2.6.6 \\
    \hline
    $R_\mathrm{g}$  & 5 May 2020 & v2.6.6 \\
    \hline
    LD1 & 23 Jun 2022 & v2.8.3 \\
    \hline
    RMSD & 5 May 2020 & v2.6.6 \\
    \hline
    $Q$ & 23 Jun 2022 & v2.8.3 \\
    \hline
    $R_\mathrm{ee}$/$R_\mathrm{g}$ (2D) & 23 Jun 2022 & v2.8.3 \\
    \hline
\end{tabular}
\caption{The version of each software used in each simulation set.}
\label{tab:software_versions}
\end{table}

\begin{table}
\begin{tabular}{|c|c|c|c|c|c|}
    \hline
    Simulations & WT HEIGHT & WT SIGMA & WT BIASFACTOR & UT HEIGHT & UT SIGMA  \\
    \hline
    1D & 1.0 $k_BT$ & 0.5 & 2.0 & --- & --- \\
    \hline
    $R_\mathrm{ee}$ & 0.4 kJ/mol & 0.06 nm & 10.0 & 0.2 kJ/mol & 0.01 nm \\
    \hline
    $R_\mathrm{g}$ & 0.4 kJ/mol & 0.02 nm & 10.0 & 0.2 kJ/mol & 0.01 nm \\
    \hline
    LD1 & 1.0 kJ/mol & 0.5 & 6.0 & --- & --- \\
    \hline
    RMSD & 0.4 kJ/mol & 0.02 nm & 10.0 & 0.2 kJ/mol & 0.01 nm \\
    \hline
    $Q$ & 0.6 kJ/mol & 0.02 & 10.0 & 0.2 kJ/mol & 0.01 \\
    \hline
    $R_\mathrm{ee}$/$R_\mathrm{g}$ (2D) & 0.6 kJ/mol & 0.06 nm/0.02 nm & 10.0 & --- & --- \\
    \hline
\end{tabular}
\caption{The MetaD parameters used for the well-tempered and untempered MetaD simulations for each CV.}
\label{tab:metad_params}
\end{table}

\end{document}